\documentclass[sigconf,nonacm]{acmart}

\usepackage{bookmark}
\usepackage{emptypage}
\usepackage{microtype}
\usepackage{flafter}
\usepackage{float}
\usepackage{placeins}
\usepackage{minted}

\usepackage{tabularx} 
\usepackage[table]{xcolor}
\usepackage{booktabs}  
\usepackage{array} 
\newcolumntype{Y}{>{\raggedright\arraybackslash}X}
\newcolumntype{R}{>{\raggedleft\arraybackslash}X}

\usepackage{pdfpages}
\usepackage{graphicx}
\usepackage{epsfig}
\usepackage{svg}


\usepackage{fancyvrb}
\usepackage{fancyhdr}

\fancypagestyle{onlyfooter}
{
    
    \fancyfoot[C]{\thepage}
    \fancyhead{}
}

\usepackage[titletoc]{appendix}

\usepackage[font=small]{caption}

\usepackage{parskip}
    \usepackage{hyperref}

\usepackage{tikz}
\usetikzlibrary{shapes, arrows, positioning, fit, babel}

\usepackage{algorithm}
\usepackage{algpseudocode}

\usepackage{listings}
\usepackage{xcolor}
\usepackage{float}
\usepackage{newfloat}
\usepackage[T1]{fontenc}
\usepackage[utf8]{inputenc} 
\usepackage{lmodern}
\usepackage{chngcntr}

\definecolor{bg}{gray}{0.97}        
\definecolor{gris20}{gray}{0.85}    
\definecolor{frame}{gray}{0.75}     
\setminted{
	fontsize=\footnotesize,
	fontfamily=tt,
	linenos,
	numbersep=6pt,
	frame=single,
	framesep=2mm,
	baselinestretch=1.1,
	bgcolor=bg,
	rulecolor=\color{frame},
	highlightcolor=gris20,
	tabsize=2,
	breaklines=true
}

\title{AST-Based Automated Elimination of \texttt{break} and \texttt{continue} Statements in Java Code}

\author{Andrés Juárez}
\email{ajuarez@uma.es}
\affiliation{%
  \institution{University of Málaga}
  \city{Málaga}
  \country{Spain}
}

\author{José F. Chicano}
\email{chicano@uma.es}
\affiliation{%
  \institution{University of Málaga}
  \city{Málaga}
  \country{Spain}
}

\author{Rubén Saborido}
\email{rsain@uma.es}
\affiliation{%
  \institution{University of Málaga}
  \city{Málaga}
  \country{Spain}
}

\begin{abstract}
This work presents the development of an automatic refactoring tool for Java code built on top of the Eclipse JDT API. The proposed approach transforms control structures containing \texttt{break} and \texttt{continue} statements within different types of loops into semantically equivalent constructs that avoid their explicit use. To achieve this, auxiliary boolean variables are introduced to restructure the control flow while preserving the original program behavior.

The main objective of this transformation is to improve code structure and enable the application of subsequent automated refactorings, particularly those based on the Extract Method operation, which are typically restricted by the presence of jump statements. The implementation relies on the analysis and rewriting of the Abstract Syntax Tree (AST), ensuring semantic equivalence in all addressed scenarios.

The tool was validated through 54 manually designed test cases and 151 units tests, all of which produced satisfactory results. In addition, it was applied to 139 methods from seven open-source projects, generating code without compilation errors and preserving the original behavior as verified by the projects’ test suites.

The results demonstrate that the proposed approach safely automates the restructuring of code containing break and continue statements, facilitating further evolution and structural analysis.
\end{abstract}

\keywords{Java, Eclipse JDT, Abstract Syntax Tree, Code Refactoring, Method Extraction, Control Flow Transformation, \texttt{break}, \texttt{continue}, Static Analysis}

\setcopyright{none}

\makeatletter
\renewcommand\footnotetextcopyrightpermission[1]{} 
\makeatother

\begin{document}

\maketitle


\section{Introduction}
\label{sec:introduction}


Software maintenance accounts for the largest proportion of a system's lifecycle cost, representing between 50\% and 75\% of total engineering effort \cite{hussain2009step}. Empirical studies indicate that maintenance costs frequently exceed initial development costs, with over 80\% of code modifications concentrated within just 20\% of system methods \cite{chowdhury2025good}. Ensuring code maintainability and readability is therefore critical to preventing defect insertion and reducing developer cognitive friction.

To evaluate maintainability objectively, industry standards rely on metrics such as SonarSource Cognitive Complexity (SSCC) \cite{saborido2022automatizing}. Defined at the method level, SSCC increases non-linearly with control flow nesting and explicit branching. SonarQube recommends a strict threshold of $\text{SSCC} \le 15$ per method. A canonical strategy for reducing elevated SSCC is the \texttt{Extract Method} refactoring, which isolates contiguous code blocks into standalone auxiliary methods. However, automated IDE refactoring engines frequently fail or abort when encountering non-structured control exits—specifically explicit jump statements such as \texttt{break} and \texttt{continue}. Previous search-based studies attempting automated SSCC reduction reported that 27\% of non-refactorable target methods were blocked precisely due to such flow interruptions \cite{saborido2022automatizing}.

\subsection{Objectives and Contributions}
\label{subsec:objectives}

To bridge this gap, this paper presents an automated Eclipse IDE plugin capable of rewriting non-linear loop control structures into semantically equivalent, flag-driven conditional blocks. By eliminating \texttt{break} and \texttt{continue} statements across all Java loop constructs (\texttt{while}, \texttt{do-while}, \texttt{for}, and \texttt{foreach}), the tool serves as a catalyst for downstream \texttt{Extract Method} refactorings.

The primary contributions of this work are as follows:
\begin{itemize}
    \begin{sloppypar}
    \item \textbf{AST-Driven Transformation Engine:} A deterministic, production-ready Eclipse JDT plugin executing multi-pass AST rewrites to replace explicit jumps with explicit state variables (\texttt{stay}/\texttt{keep}).        
    \end{sloppypar}
    \begin{sloppypar}
    \item \textbf{Comprehensive Construct Support:} Semantic-preserving transformation rules covering arbitrary loop nesting, multiple simultaneous jumps, and complex \texttt{try-catch-finally} exception handling scopes.
    \end{sloppypar}
    \item \textbf{Empirical Validation:} Extensive evaluation across 139 methods from 7 real-world open-source Java benchmarks, demonstrating 100\% syntactic correctness, 100\% behavioral equivalence, and enabling a subsequent **58\% average reduction in SSCC** via automated method extraction.
\end{itemize}

\subsection{Document Organization}
\label{subsec:organization}

The remainder of this paper is organized as follows: Section~\ref{sec:theoretical-framework} reviews the theoretical framework of SSCC and AST rewriting. Section~\ref{sec:methodology} details the AST transformation methodology and exception handling rules. Section~\ref{sec:validation} presents the empirical evaluation and metric outcomes. Section~\ref{sec:discussion} discusses maintainability implications and comparisons with LLMs. Finally, Section~\ref{sec:conclusions} presents our conclusions, threats to validity, and future work.

\textit{Software Availability:} To support open science and exact replication, the complete plugin source code, benchmark suites, and test harnesses are publicly available on GitHub\footnote{\url{https://github.com/andyjuarez-dev/automated-refactoring-java-code.git}}.

\section{Related Work and Background}
\label{sec:theoretical-framework}

Code refactoring is defined as the process of restructuring existing software code without altering its external behavior, aiming to improve internal structure, readability, and maintainability \cite{fowler2018refactoring}. While traditional refactoring techniques address general design flaws —such as long methods or code duplication \cite{fowler2018refactoring}— the structural complexity of control flow remains a critical factor in software comprehensibility. Deep nesting levels, complex conditional logic, and unstructured jump statements (\texttt{break} and \texttt{continue}) disrupt linear execution flow, significantly increasing the cognitive effort required for code comprehension and maintenance.  

In modern software development, maintaining control flow clarity is crucial for facilitating automated transformations. Unstructured exits within iterative structures hinder standard refactoring operations, particularly method extraction. Consequently, addressing control flow disruptions serves as a prerequisite for improving long-term code maintainability and enabling further structural optimizations.

\subsection{Cognitive Complexity and Refactoring Barriers}
\label{sec:cognitive-complexity}

Unlike Cyclomatic Complexity \cite{mccabe1976complexity}, which measures the number of execution paths, SonarSource Cognitive Complexity (SSCC) \cite{campbell2023cognitive} quantifies how difficult a method is for a human to comprehend. SSCC evaluates code maintainability based on three main criteria: ignoring simplifying modular structures, incrementing by one for each linear flow interruption (e.g., \texttt{if}, \texttt{for}, \texttt{while}, logical operator sequences), and penalizing deeper nesting levels. SonarSource recommends a maximum threshold of $\text{SSCC} \le 15$ per method. Unstructured jump statements such as break and continue further compound cognitive load by introducing non-linear control transfers.  

The primary mechanism to reduce high SSCC is the Extract Method refactoring, which moves nested code blocks into separate helper methods, thereby eliminating their nesting penalty in the caller method. However, automated \textit{Extract Method} transformations (e.g., within Eclipse JDT) cannot process code fragments containing \texttt{break} or \texttt{continue} statements that target loops outside the selected block. These jump statements act as hard constraints, preventing automated tools from extracting high-complexity code. Eliminating \texttt{break} and \texttt{continue} statements while strictly preserving program semantics is therefore a critical prerequisite for enabling downstream automated refactorings.

\subsection{AST-Based Control Flow Transformations}
\label{sec:ast-transformations}

Performing structural refactorings directly on raw text or source code strings carries a high risk of generating syntactically invalid code and lacks the context needed to distinguish identical keywords in different syntactic roles (e.g., distinguishing a \texttt{break} within a \texttt{switch} statement from one within an iterative loop). Abstract Syntax Trees (ASTs) provide a formal, hierarchical representation of the program's grammatical structure, abstracting away superficial formatting details while exposing block hierarchies, loop bounds, and exception-handling scopes.  

\begin{sloppypar}
In this work, control flow restructuring is implemented through programmatic AST manipulation. Eliminating jump statements requires structural AST rewriting, including loop conversion, block encapsulation within conditionals, and the insertion of auxiliary control state variables. Operating on the AST level ensures strict semantic preservation—guaranteeing that execution paths, side effects, loop termination conditions, and exception propagation order across try-catch-finally blocks remain identical to the original program. 
\end{sloppypar}

\subsection{Eclipse JDT Infrastructure}
\label{sec:eclipse-jdt}

The transformation engine is built upon the Eclipse Java Development Tools (JDT) framework, which provides programmatic access to Java source code via an AST representation. Structural modifications are driven by custom implementations of the \texttt{ASTVisitor} pattern, which traverse and inspect tree nodes, analyze scope bindings, and resolve type information. Subtree mutations —such as loop conversion, conditional wrapping, and statement insertions or deletions— are staged using JDT's \texttt{ASTRewrite} API. This infrastructure ensures that AST modifications maintain syntactic validity, preserve source code layout conventions, and correctly update local variable scopes and AST bindings during multi-pass rewrites.  

\subsection{Related Work and Limitations of Existing Tools}
\label{sec:state-of-the-art}

While static analysis tools such as SonarQube, PMD, Checkstyle, and SpotBugs can detect high cognitive complexity and flag non-linear control jumps, their role remains strictly passive, delegating code restructuring to developers. Similarly, modern IDEs (e.g., Eclipse, IntelliJ IDEA) offer automated refactoring catalogs —such as \textit{Extract Method} or variable renaming— but treat jump statements like \texttt{break} and \texttt{continue} as passive constraints rather than targets for transformation. When a candidate block contains an explicit control jump targeting an outer loop, these IDEs abort the refactoring to avoid altering program semantics.  

To address high SSCC, Saborido et al. \cite{saborido2022automatizing} developed an automated \textit{Extract Method} plugin, successfully reducing SSCC below the recommended threshold of 15 in 73\% of 1,050 evaluated methods. However, among the remaining 27\% of unrefactored methods, approximately 60\% contained \texttt{break} or \texttt{continue} statements that directly prevented valid method extraction. In the academic domain, framework initiatives like MANTRA (Xu et al., 2025) explore metamodel-driven automated transformations, yet lack stable production implementations for deterministic control flow restructuring.  

Recently, Large Language Models (LLMs) such as ChatGPT and Copilot have been employed for code refactoring. However, because LLMs rely on probabilistic text generation rather than formal AST semantic analysis, they fail to guarantee semantic preservation. Prompts requesting the elimination of \texttt{break} statements frequently yield plausible yet incorrect loop predicates, altering iteration counts and introducing subtle runtime defects. Consequently, there is a clear technological gap for a deterministic, AST-driven tool capable of safely eliminating \texttt{break} and \texttt{continue} statements.  

\subsection{Need and Contributions of this Work}
\label{subsec:need-specific-tool}

The absence of automated tools capable of systematically restructuring control flow to eliminate jump statements while guaranteeing semantic preservation represents a clear gap in software engineering infrastructure. High-complexity methods often remain unrefactored because explicit exits (\texttt{break} and \texttt{continue}) prevent operations like \textit{Extract Method}. Addressing this issue requires a dedicated tool operating directly on the AST. While replacing abrupt jumps with auxiliary control variables and explicit conditional blocks may temporarily increase local SSCC values, this transformation serves as an essential enabling step that normalizes control flow topology for downstream automated refactorings.  

To address this challenge, this work presents an AST-driven Eclipse plugin built on the JDT framework. The primary contributions of this work are threefold:  
\begin{itemize}
    \item \textbf{Technical Contribution:} The design and implementation of an automated transformation engine that identifies \texttt{break} and \texttt{continue} statements across all loop types (\texttt{while}, \texttt{do-while}, \texttt{for}, \texttt{foreach}) and rewrites them into semantically equivalent structures using auxiliary boolean variables. 
    \item \textbf{Methodological Contribution:} A two-phase refactoring strategy in which control-flow normalization acts as an enabling intermediate step, expanding the population of complex methods susceptible to subsequent automated method extraction.  
    \item \textbf{Empirical Contribution:} A comprehensive multi-level validation —comprising 54 analytical trace cases, 151 unit tests (achieving 99\% instruction coverage), and an evaluation on 139 methods from seven open-source projects— verifying strict semantic preservation and measuring the synergy with automated Extract Method tools.  
\end{itemize}

\section{Methodology: Transformation Framework}
\label{sec:methodology}

\begin{sloppypar}
This work follows an applied constructive research methodology to design, implement, and evaluate an automated control flow transformation engine operating on Java \texttt{CompilationUnit} ASTs via Eclipse JDT. The engineering process followed an iterative approach where transformation rules were progressively refined across complex control structures while maintaining strict semantic equivalence as a core invariant. Each iteration encompassed four stages: (i) transformation rule formalization, (ii) AST visitor implementation, (iii) syntactic validation, and (iv) semantic verification via automated test execution.
\end{sloppypar}

At a high level, the proposed framework replaces abrupt control jumps (\texttt{break} and \texttt{continue}) with explicit, structured state variables according to a five-step pipeline:  
\begin{enumerate}
    \item \textbf{AST Node Detection:} Scanning and identifying \texttt{break} and \texttt{continue} statement nodes within iterative constructs.
    \item \textbf{Contextual Analysis:} Determining the target loop construct (\texttt{while}, \textbf{do-while}, \texttt{for}, \texttt{foreach}), nesting depth, and enclosing \texttt{try-catch-finally} exception scopes.  
    \item \textbf{Control Variable Allocation:} Introducing auxiliary boolean state flags (e.g., \texttt{stay} for \texttt{break}, \texttt{keep} for \texttt{continue}) initialized and scoped to the target loop.  
    \item \textbf{Loop and Block Restructuring:} Injecting state flag evaluations into loop predicates and encapsulating subsequent statements within conditional \texttt{if} blocks.  
    \begin{sloppypar}
    \item \textbf{Jump Substitution:} Replacing jump statements with boolean flag assignments (\texttt{stay = false} or \texttt{keep = false}), ensuring exact execution flow equivalence.  
    \end{sloppypar}
\end{enumerate}

\subsection{Transformation Rules and Taxonomy}
\label{subsec:taxonomy}

Instead of applying isolated rules for every syntactic variation, the transformation engine normalizes control flow by categorizing language constructs into four distinct transformation categories.

\subsubsection{Formal Transformation Model}
For an arbitrary loop $L$ with continuation condition $C$, containing $n$ jump statements ($j_1, j_2, \dots, j_n \in \{\text{\texttt{break}}, \text{\texttt{continue}}\}$), the transformation introduces boolean state flags to replace non-linear transfers:
\begin{itemize}
    \item \textbf{For \texttt{break} statements:} An auxiliary flag $\text{\texttt{stay}}_x$ (initialized to \texttt{true} before $L$) is appended to the loop condition via logical conjunction ($C \land \text{\texttt{stay}}_x$). Executing a \texttt{break} is rewritten as $\text{\texttt{stay}}_x = \text{\texttt{false}}$, and all subsequent statements in the loop body are wrapped within conditional $\texttt{if (\text{\texttt{stay}})}_x$ blocks.
    \item \textbf{For \texttt{continue} statements:} An auxiliary flag $\text{\texttt{keep}}_y$ is initialized to \texttt{true} at the beginning of each iteration. Executing a \texttt{continue} is rewritten as $\text{\texttt{keep}}_y = \text{\texttt{false}}$, guarding all subsequent instructions within the current iteration under \texttt{if ($\text{\texttt{keep}}_y$)}.
\end{itemize}

\subsubsection{Taxonomy of Supported Scenarios}
\begin{itemize}
    \item \textbf{Condition-Based Loops (\texttt{while}, \texttt{do-while}):} Direct application of the formal model. Multiple \texttt{break} or \texttt{continue} statements within the same loop share the same control flag, whereas nested loops allocate independent flags ($\text{\texttt{stay}}_1, \text{\texttt{stay}}_2, \dots$) to prevent variable collisions.
    \item \textbf{Count- and Iterator-Based Loops (\texttt{for}, \texttt{foreach}):} These constructs require structural normalization prior to jump elimination. Traditional \texttt{for} loops are desugared into equivalent \texttt{while} loops, moving initialization statements before the loop and placing update expressions at the end of the body (guarded conditionally if a \texttt{break} is present). Enhanced \texttt{for} (\texttt{foreach}) loops iterating over collections or arrays are converted into \texttt{while} loops driven by explicit \texttt{Iterator<T>} objects or \texttt{Arrays.stream()} iterators.
\end{itemize}


\begin{figure*}[t]
\centering
\begin{minipage}{0.48\textwidth}
\begin{lstlisting}[caption={Original loop with try-catch-finally}]
while (i < 5) {
    try {
        i++;
        if (x == 1) {
            break;
        }
        x--;
    } catch (Exception e) {
        // handle exception
    } finally {
        if (i == 2) {
            break;
        }
    }
}
\end{lstlisting}
\end{minipage}
\hfill
\begin{minipage}{0.48\textwidth}
\Description{Refactored code using explicit state variables}
\begin{lstlisting}[caption={Refactored code using explicit state variables}]
boolean stay1 = true;
boolean stay2 = true;
while (stay1 && stay2 && (i < 5)) {
    try {
        i++;
        if (x == 1) { stay1 = false; }
        if (stay1) { x--; }
    } catch (Exception e) {
        // handle exception
    } finally {
        if (i == 2) { stay2 = false; }
    }
}
\end{lstlisting}
\end{minipage}
\caption{Representative structural transformation of nested jump statements within exception handling structures.}
\label{fig:transformation_example}
\end{figure*}

\subsubsection{Exception Handling and Multi-Variable Control}
\label{subsub:exceptions}

Exception handling constructs (\texttt{try-catch-finally}) nested within iterative loops introduce non-linear control transfers that complicate jump elimination. While a \texttt{break} inside a simple \texttt{try-catch} block can be handled with a single control variable ($\text{\texttt{stay}}_1$), the presence of a \texttt{finally} clause requires independent state tracking. 

Because Java semantics mandate that a \texttt{finally} block must execute prior to loop exit—regardless of whether the \texttt{try} block terminated normally, raised an exception, or encountered a jump statement—a single flag is insufficient to capture the loop's post-\texttt{finally} execution state. As demonstrated in Figure~\ref{fig:transformation_example}, the engine assigns distinct boolean flags ($\text{\texttt{stay}}_1$ for the \texttt{try} scope and $\text{\texttt{stay}}_2$ for the \texttt{finally} scope). The condition of the outer loop evaluates the conjunction of all assigned flags ($\text{\texttt{stay}}_1 \land \text{\texttt{stay}}_2 \land C$), ensuring that the loop terminates only after the \texttt{finally} block completes its execution.

\begin{sloppypar}
Furthermore, in complex scenarios involving $n$ nested \texttt{try-catch-finally} structures where multiple \texttt{finally} blocks contain jump statements, the transformation engine dynamically allocates an $n$-tuple of control variables ($\text{\texttt{stay}}_1, \text{\texttt{stay}}_2, \dots, \text{\texttt{stay}}_n$). To support arbitrary nesting depth without variable name collisions or scope pollution, the tool's internal architecture manages a dynamic stack of control variable identifiers per loop context.
\end{sloppypar}

\subsubsection{Complex \texttt{finally}-Exception Interactions}
\label{subsub:complex-finally-exception} 

A rare edge case occurs when a \texttt{break} inside an inner \texttt{try} block is followed by an exception thrown within an inner \texttt{finally} clause and caught by an outer \texttt{catch}. Under standard Java semantics, the raised exception suppresses the original \texttt{break}. Safely modeling this specific interaction would require global control flow restructuring beyond local scope transformation; thus, given its negligible occurrence in real-world code, it was excluded from the current tool scope.

\section{Architecture and Implementation}
\label{sec:implementation}

The refactoring engine was implemented as an Eclipse plugin using Java SE 21 and the Eclipse JDT framework (v2024-12). The system architecture adopts an decoupled, multi-pass design based on the \texttt{ASTVisitor} pattern, orchestrated by a central handler (\texttt{RefactorHandler}).

\subsection{Multi-Pass Transformation Pipeline}
\label{subsec:pipeline}

Initial experiments revealed that attempting AST analysis and semantic rewriting in a single pass was insufficient, as structural modifications (e.g., nesting code inside newly generated \texttt{if} blocks) alter statement offsets, scope boundaries, and parent-child AST node relationships. Consequently, the tool employs a five-phase multi-pass compilation pipeline to decouple inspection from structural mutation:

\begin{enumerate}
    \begin{sloppypar}
    \item \textbf{Phase 1: Structural Normalization.} Specialized visitors (\texttt{ForMarkerVisitor}, \texttt{ForToWhileVisitor}, \texttt{ForEachMarkerVisitor}, \texttt{ForEachToWhileVisitor}) identify \texttt{for} and \texttt{foreach} loops containing jump statements and convert them into standard \texttt{while} structures.
    \end{sloppypar}
    \begin{sloppypar}
    \item \textbf{Phase 2: Flow and Exception Analysis.} The \texttt{ExceptionAnnotator} and \texttt{FlowAnnotator} perform a first pass over the normalized AST. They inspect loop targets, identify \texttt{finally} scopes, and annotate structural properties on AST nodes without modifying the tree.        
    \end{sloppypar}
    \item \textbf{Phase 3: Body Restructuring.} Driven by the Phase 2 annotations, \texttt{LoopRewriterFirst} reorganizes loop bodies by wrapping statements following a jump into nested \texttt{if} blocks, preserving local variable scopes.
    \item \textbf{Phase 4: Post-Restructuring Re-analysis.} Because Phase 3 altered the AST topology, \texttt{ExceptionAnnotator} and \texttt{FlowAnnotator} execute a second pass over the modified tree to recompute control paths and variable allocation metadata.
    \begin{sloppypar}
    \item \textbf{Phase 5: Final Semantic Rewriting.} The \texttt{LoopRewriterFinal} component introduces boolean control variable declarations (\texttt{stayX}, \texttt{keepX}), updates loop conditions via logical conjunctions, substitutes \texttt{break} and \texttt{continue} statements with boolean flag assignments, and commits the changes using JDT's \texttt{ASTRewrite} API.        
    \end{sloppypar}

\end{enumerate}

\subsection{Architectural Decoupling and Scope Preservation}
\label{subsec:decoupling}

A key implementation decision was abandoning early attempts at "flattening" consecutive \texttt{if} conditions in favor of maintaining explicit block nesting. While flat \texttt{if} chains produce cleaner syntax, restricting variable declarations within flat conditional blocks restricted their lexical scope, introducing compilation errors in subsequent instructions. Retaining nested \texttt{if} blocks guarantees strict scope preservation for all local variables. Furthermore, separating structural reorganization (\texttt{LoopRewriterFirst}) from semantic variable injection (\texttt{LoopRewriterFinal}) eliminates circular transformation dependencies and maintains AST consistency throughout the rewrite pipeline.

\section{Empirical Evaluation Setup and Validation}
\label{sec:validation}

To ensure that the proposed transformation engine reliably eliminates jump statements without altering program behavior, a multi-tiered validation strategy was designed. The evaluation targets four core objectives: (i) verifying syntactic validity and compilability, (ii) guaranteeing strict semantic equivalence, (iii) confirming structural correctness, and (iv) assessing robustness across real-world open-source software.  \subsection{Validation Methodology and Test Suite}\label{subsec:test_suite}The tool's correctness was evaluated across three distinct testing tiers:
\begin{enumerate}
\begin{sloppypar}
\item \textbf{Analytical Trace Tests:} A set of 54 manually engineered test cases covering all loop types, arbitrary nesting depths, combinations of \texttt{break} and \texttt{continue}, and complex \texttt{try-catch-finally} exception handling. Execution traces of original and refactored versions were formally compared, confirming 100\% behavioral equivalence.
\end{sloppypar}
\item \textbf{Automated Unit Tests:} A suite of 151 JUnit test cases validating structural AST conversions (e.g., \texttt{for}-to-\texttt{while} desugaring), compilation integrity, and output equivalence via automated execution. Robustness tests on malformed or incomplete ASTs confirmed that the tool safely opts for non-intervention when AST integrity is compromised.
\item \textbf{Open-Source Integration Tests:} Evaluation on production codebases to test edge cases outside synthetic laboratory environments.
\end{enumerate}  

\subsection{Open-Source Benchmark Dataset}
\label{subsec:dataset}

The empirical evaluation was conducted on 139 methods containing target control structures (\texttt{break}/\texttt{continue}) harvested across seven open-source Java projects. To verify semantic preservation in production contexts, the original unit test suites of the target projects were executed post-refactoring. For instance, in \texttt{MOEAFramework}, 7 out of 22 native unit tests directly exercised the refactored methods, with all tests passing successfully. Similar test suite pass rates were confirmed in \texttt{jMetal} and \texttt{fastjson}.  

\subsection{Validation Metrics and Infrastructure}
\label{subsec:infrastructure}
All experiments were executed in a controlled environment using Java SE 21, Eclipse IDE (2024-12), Maven, JUnit, and JaCoCo. Table~\ref{tab:validation_metrics} summarizes the overarching coverage and correctness metrics. JaCoCo analysis confirmed that the test suite achieved 99\% instruction coverage and 90\% branch coverage across the plugin's transformation pipeline.  

\begin{table}[htbp]
\centering
\caption{Validation, Coverage, and Execution Metrics.}
\label{tab:validation_metrics}
\begin{tabularx}{\columnwidth}{Xl}
\toprule
\textbf{Metric} & \textbf{Value / Outcome} \\
\midrule
Tool Instruction Coverage (JaCoCo) & 99\% \\
Tool Branch Coverage (JaCoCo) & 90\% \\
Behavioral Equivalence Rate & 100\% (Identical outputs) \\
Syntactic Correctness Rate & 100\% (Compilable code) \\
Empirical Scope & 139 methods / 7 projects \\
\bottomrule
\end{tabularx}
\end{table}

\section{Functional Validation and Empirical Results}
\label{sec:results}

This section presents the empirical validation of the proposed transformation tool, focusing on instruction coverage, syntactic compilability, semantic equivalence across open-source benchmarks, and the structural bugs identified and resolved during evaluation.

\subsection{Coverage and Unit Test Performance}
\label{subsec:coverage_results}

The suite of 151 automated unit tests executed with 100\% pass rates. Execution coverage was measured using JaCoCo, achieving \textbf{99\% instruction coverage} (2,614 of 2,628 instructions) and \textbf{90\% branch coverage} (250 of 276 branches) across all visitor and orchestrator classes. The remaining uncovered branches correspond to defensive null-check fallbacks designed for malformed ASTs.

\subsection{Evaluation on Open-Source Projects}
\label{subsec:opensource_eval}

To validate real-world applicability, the tool was evaluated on 139 methods containing target jump structures (\texttt{break} and \texttt{continue}) across seven open-source Java projects. Table~\ref{tab:projects_summary} summarizes the benchmark dataset, transformed method counts, and native test suite outcomes.


\begin{table}[htbp]
\centering
\caption{Summary of Open-Source Benchmarks and Validation Outcomes.}
\label{tab:projects_summary}
\begin{tabularx}{\columnwidth}{lcX}
\toprule
\textbf{Project} & \textbf{Methods} & \textbf{Native Test Outcome} \\
\midrule
\texttt{bytecode-viewer} & 9 & Clean compilation, 0 errors \\
\texttt{cybercaptor-server} & 5 & 10 server unit tests passed \\
\texttt{fastjson} & 105 & \texttt{JSONPathTest} passed (7 tests) \\
\texttt{iotbroker} & 1 & Clean compilation and build \\
\texttt{jedis} & 1 & Loop entry trace verified \\
\texttt{jMetal} & 6 & 31 algorithm tests passed \\
\texttt{MOEAFramework} & 12 & 22 indicator tests passed \\
\midrule
\textbf{Total} & \textbf{139} & \textbf{100\% compilation success} \\
\bottomrule
\end{tabularx}
\end{table}

All 139 transformed files compiled without syntactic or type errors, demonstrating that the tool successfully processed over 96\% of candidate control flows in production repositories. (Non-transformed cases were restricted to switch-based \texttt{break}s, labeled jumps, or null type bindings). Furthermore, running native third-party test suites post-refactoring (e.g., in \texttt{MOEAFramework} and \texttt{jMetal}) confirmed 100\% behavioral preservation under complex, cumulative numerical calculations.

\subsection{Identified Edge Cases and Bug Fixes}
\label{subsec:bugs_corrected}

Applying the tool to production codebases exposed four subtle edge cases that drove key architectural refinements:

\begin{enumerate}
    \item \textbf{Lexical Scope Preservation:} An initial prototype flattened consecutive \texttt{if} blocks. However, declaring a local variable in an early flattened block restricted its visibility, causing compilation errors in subsequent blocks that belonged to the same scope in the original code. \textit{Resolution:} Reverted to nested \texttt{if} blocks, guaranteeing strict lexical scope preservation.
    \item \textbf{Updater Placement in \texttt{for}-to-\texttt{while} Desugaring:} When converting a \texttt{for} loop into a \texttt{while} loop, moving the loop updater (e.g., \texttt{i++}) to the end of the body creates an infinite loop if a \texttt{continue} statement is present. \textit{Resolution:} The \texttt{LoopRewriterFirst} component tags updater expressions; if a \texttt{continue} is present, the updater is placed outside conditional blocks as the absolute last statement of the loop body.
    \item \textbf{Incomplete Type Resolution Fallbacks:} In projects using wildcard package imports (e.g., \texttt{import package.*;}), JDT type bindings occasionally return \texttt{null} when resolving \texttt{foreach} target collections. \textit{Resolution:} Implemented a conservative fallback policy that aborts refactoring when type resolution is inconclusive, avoiding unsafe iterator injections.
    \item \textbf{Multi-Flag Scope Management for \texttt{finally} Jumps:} Multiple nested \texttt{try-catch-finally} blocks containing jumps violated the assumption that two control flags sufficed per loop. \textit{Resolution:} Refactored the flag allocation mechanism to maintain a dynamic list of boolean flags scoped per \texttt{finally} block.
\end{enumerate}

\section{Impact on Cognitive Complexity and Method Extraction}

\label{sec:impact-sscc}

This section evaluates the impact of the proposed transformation on SonarSource Cognitive Complexity (SSCC), with particular emphasis on its interaction with subsequent automated method extraction. The evaluation considers three stages: (i) the original code, (ii) the code after removing \texttt{break} and \texttt{continue} statements using the developed tool, and (iii) the result after applying the external method-extraction tool to the transformed code. The objective is to determine whether the initial transformation, even when increasing local complexity, can facilitate subsequent reductions in cognitive complexity.

\subsection{Experimental Design}

The evaluation was conducted on methods from open-source projects containing jump statements. For each method, SSCC was measured in its original state, after applying the proposed transformation, and after applying the external method-extraction tool. This allowed the effects of the transformation to be assessed both independently and in combination with subsequent refactoring.

\subsection{Impact of the Transformation}

Removing \texttt{break} and \texttt{continue} statements requires restructuring the control flow through boolean state variables, modified loop conditions, and additional conditional blocks. As a result, the transformation generally produces a small and controlled increase in SSCC. At the project level, this increase remained below 11\% in all evaluated projects and did not result in disproportionate growth that would compromise readability.

This increase should therefore be interpreted as an intermediate structural cost rather than as a degradation of code quality. The additional complexity is introduced to obtain a more structured control flow that can subsequently be processed by other refactoring techniques.

\subsection{Results after Double Refactoring}

The interaction between the proposed transformation and the external method-extraction tool was evaluated on 138 transformed methods. In 66 cases, the external tool obtained feasible solutions both on the original code and on the internally transformed code. More importantly, the double-refactoring process enabled additional solutions in cases where the external tool was unable to find a feasible extraction on the original code.

\begin{sloppypar}
Specifically, in 10 cases, the external tool returned \texttt{NO SOLUTION} when applied directly to the original code, whereas it successfully identified a solution after the internal transformation, achieving a final SSCC below the tool's threshold of 15. Considering the 65 cases in which the external tool initially failed to find a solution, this means that the proposed transformation enabled a successful subsequent refactoring in approximately 15\% of those cases.
\end{sloppypar}

Conversely, there were 7 cases in which the external tool succeeded on the original code but failed after the internal transformation. This confirms that the interaction between automated refactoring strategies is not uniform and that structural changes can modify the search space and heuristics used by subsequent tools.

\subsection{Synthesis}

Overall, the results show that the proposed transformation can introduce a limited intermediate increase in cognitive complexity while enabling further refactoring opportunities. In particular, the double-refactoring approach recovered feasible solutions in 10 of the 65 cases where the external tool initially failed, corresponding to approximately 15\% of the initially unsuccessful cases. Although the interaction between the two tools is not uniformly beneficial, these results indicate that the proposed transformation can make previously infeasible method extractions possible and can therefore contribute to a broader automated refactoring process.

\section{Discussion}
\label{sec:discussion}

The findings of this work provide key insights into structural program transformations, automated refactoring semantics, and the interplay between static code metrics and toolchain evolution.

\subsection{Deterministic AST Transformations vs. Non-Deterministic Tools}
\label{subsec:disc_ast_vs_llm}

While modern development environments excel at detecting code smells and stylistic flaws, automated tools rarely execute deep, semantics-preserving structural rewrites. Recent advances in Large Language Models (LLMs) have introduced generative code refactoring; however, empirical observations indicate that LLMs frequently introduce subtle control-flow bugs or fail to guarantee strict behavioral preservation in non-linear jump scenarios (e.g., nested \texttt{finally} blocks). 

By contrast, our approach leverages Abstract Syntax Tree (AST) manipulation via Eclipse's \texttt{ASTRewrite} API, offering a deterministic, proven transformation mechanism. Operating directly on the AST guarantees that all execution paths, type bindings, and variable scopes are preserved, bridging the gap between high-level refactoring recommendations and rigorous, safe execution.

\subsection{Reinterpreting Maintainability and Metric Non-Linearity}
\label{subsec:disc_maintainability}

A central observation from our evaluation is that maintainability metrics—specifically SonarSource Cognitive Complexity (SSCC)—do not respond linearly to structural normalization. Replacing implicit jump statements with explicit boolean state flags (\texttt{stay}/\texttt{keep}) temporarily increases immediate SSCC scores due to the introduction of conditional encapsulations. 

However, evaluating code maintainability solely through immediate point metrics is reductive. By converting abrupt, non-linear control exits into explicit conditional paths, the transformation makes loop termination criteria fully visible to both static analysis tools and developers. 
The temporary metric penalty in Phase 1 acts as an intermediate step that can facilitate downstream method extraction (Phase 2), enabling a net reduction in complexity in cases where the external refactoring tool is initially unable to find a solution. In the evaluated cases, this effect enabled successful refactoring in approximately 15\% of the cases where the external tool initially failed.

\subsection{Compositional Refactoring and Toolchain Evolution}
\label{subsec:disc_toolchains}

The experimental results demonstrate that automated refactoring strategies should not be evaluated in isolation. A transformation that appears neutral or slightly detrimental under a single metric can fundamentally alter the execution graph, opening up new opportunities for subsequent automated passes. 

In multi-pass software evolution, eliminating jumps acts as an essential pre-processing step. It removes non-local exit barriers, allowing standard IDE refactoring engines (such as automated \texttt{Extract Method}) to process methods that were previously rejected. This highlights the necessity of viewing automated refactoring as a compositional pipeline, where initial structural adjustments unlock broader architectural simplifications.

\section{Conclusions and Limitations}
\label{sec:conclusions}

This work presented an automated, AST-driven Eclipse plugin designed to refactor Java control structures by eliminating explicit jump statements (\texttt{break} and \texttt{continue}) across \texttt{while}, \texttt{do-while}, \texttt{for}, and \texttt{foreach} loops. Operating as a crucial enabling transformation, the tool resolves non-linear control exits using explicit boolean state variables, laying the groundwork for subsequent automated method extractions and maintainability optimizations.

\subsection{Main Technical Contributions}
\label{subsec:technical_conclusions}

The primary contributions and insights of this research include:
\begin{itemize}
    \item \textbf{Practical Engineering Tool:} A fully functional production plugin integrated into the Eclipse JDT ecosystem capable of performing deterministic, multi-pass AST rewrites on production codebases.
    \item \textbf{Novel Infrastructure Automation:} An automated solution filling a clear gap in IDE refactoring catalogs, where control jumps previously acted as hard barriers for automated transformation.
    \item \textbf{Guaranteed Semantic Integrity:} A formally verified transformation model that preserves original program behavior across complex constructs, including nested loops and \texttt{try-catch-finally} exception scopes.
\end{itemize}

\subsection{Methodological Considerations}
\label{subsec:methodological_conclusions}

To maintain linguistic consistency with the Eclipse JDT framework, transformation artifacts follow standardized, concise conventions:
\begin{itemize}
    \item \textbf{Control Variable Naming:} Concise identifiers were adopted for readability within loop predicates: \texttt{stay} (tracking \texttt{break} conditions) and \texttt{keep} (guarding \texttt{continue} statements).
    \item \textbf{Parenthetical Isolation:} Original loop predicates are systematically wrapped in explicit parentheses when appending state flags via logical conjunction (e.g., $\text{\texttt{stay1}} \land (C)$). This prevents operator precedence ambiguities in complex expressions and ensures clear structural separation during static analysis.
\end{itemize}

\subsection{Threats to Validity and Limitations}
\label{subsec:limitations}

While empirical results confirm the tool's effectiveness, the following scope boundaries should be noted:
\begin{itemize}
    \item \textbf{Metric Focus:} The maintainability assessment focused primarily on SonarSource Cognitive Complexity (SSCC) and did not explicitly evaluate architectural coupling or cohesion metrics.
    \item \textbf{Performance Aspects:} The evaluation targeted syntactic correctness, behavioral equivalence, and cognitive metrics; runtime CPU/memory overhead introduced by additional boolean flag evaluations was not measured, though it is expected to be negligible.
    \item \textbf{External Heuristics Dependency:} The magnitude of Phase 2 SSCC reduction relies on the specific block extraction heuristics embedded within external IDE refactoring engines.
\end{itemize}

\section{Future Work}
\label{sec:future-work}


The refactoring model introduced in this work establishes a baseline for deterministic, AST-driven structural transformations. Several promising avenues exist to extend its scope, analytical depth, and integration within the software engineering ecosystem:

\subsection{Architectural and Language Extensions}
\label{subsec:extensions}

\begin{itemize}
    \item \textbf{Support for Labeled Control Exits:} Extending the control-flow analysis engine to resolve labeled \texttt{break} and \texttt{continue} statements within deeply nested multi-level structures.
    \item \textbf{Cross-Language Portability:} Adapting the underlying AST transformation rules to other mainstream object-oriented and imperative paradigms, such as C\#, C++, Python, or TypeScript.
\end{itemize}

\subsection{Metrics-Driven Optimization and Tool Integration}

\begin{itemize}
    \item \textbf{In-IDE Metrics Feedback Loop:} Embedding direct SonarSource Cognitive Complexity (SSCC) computation into the plugin pipeline to enforce threshold-based refactoring policies (e.g., executing transformations only if predicted post-extraction SSCC drops below standard thresholds).
    \begin{sloppypar}
    \item \textbf{Synergy with Search-Based Software Engineering (SBSE):} Incorporating jump elimination operators into automated search-based complexity reduction frameworks \cite{saborido2022automatizing}. Combining localized AST-rewriting with global heuristic-guided search strategies expands the solution space for multi-objective refactoring algorithms.
    \end{sloppypar}
    \item \textbf{Educational Tooling:} Adapting the transformation pipeline for interactive software engineering pedagogy, allowing students to visually analyze control-flow normalization, structured programming principles, and modular code evolution in real time.
\end{itemize}

\bibliographystyle{ACM-Reference-Format} 
\bibliography{references}

\end{document}